\documentclass[a4paper,10pt, twocolumn]{article}

\usepackage[T1]{fontenc}
\usepackage[scaled=0.8]{beramono}
\usepackage[cm]{fullpage}
\usepackage{tabularx}
\usepackage{listings}
\usepackage{xcolor}
\usepackage{enumitem}
\usepackage{amsmath}
\usepackage{hyperref}
\usepackage{graphicx}
\usepackage[capitalise]{cleveref}
\usepackage{titlesec}
\usepackage{booktabs}
\usepackage{algorithm}
\usepackage[noend]{algpseudocode}
\usepackage{esvect}
\usepackage{amssymb}
\usepackage{titling}
\usepackage{lscape}
\usepackage{bold-extra}
\usepackage[font=small,labelfont=bf]{caption}
\usepackage{authblk}
\usepackage{subcaption}
\usepackage{adjustbox}
\usepackage{multirow}
\usepackage[title]{appendix}
\usepackage{wasysym}
\usepackage{tcolorbox}
\usepackage{color}
\usepackage{fancyhdr}
\algdef{SE}[DOWHILE]{Do}{doWhile}{\algorithmicdo}[1]{\algorithmicwhile\ #1}%
\algnewcommand{\LeftComment}[1]{\Statex \(\triangleright\) #1}

\makeatletter
\renewcommand{\ALG@beginalgorithmic}{\footnotesize}
\makeatother

\definecolor{lightbg}{gray}{0.95}
\definecolor{dkgreen}{rgb}{0,0.6,0}
\definecolor{gray}{rgb}{0.5,0.5,0.5}
\definecolor{mauve}{rgb}{0.58,0,0.82}

\lstdefinestyle{cppblock}{
    backgroundcolor=\color{lightbg},
    language=[GNU]C++,
    aboveskip=1mm,
    belowskip=1mm,
    showstringspaces=false,
    columns=flexible,
    basicstyle={\scriptsize\ttfamily},
    numberstyle=\emptyaccsupp,
    keywordstyle=\color{blue},
    commentstyle=\color{dkgreen},
    stringstyle=\color{mauve},
    frame=single,
    breaklines=true,
    breakatwhitespace=true,
    tabsize=4,
}

\lstdefinestyle{xsmall_cppblock}{
    backgroundcolor=\color{lightbg},
    language=[GNU]C++,
    aboveskip=1mm,
    belowskip=1mm,
    showstringspaces=false,
    columns=flexible,
    basicstyle={\tiny\ttfamily},
    numberstyle=\emptyaccsupp,
    keywordstyle=\color{blue},
    commentstyle=\color{dkgreen},
    stringstyle=\color{mauve},
    frame=single,
    breaklines=true,
    breakatwhitespace=true,
    tabsize=4,
}

\lstdefinestyle{shellblock}{
    backgroundcolor=\color{lightbg},
    language=bash,
    aboveskip=1mm,
    belowskip=1mm,
    showstringspaces=false,
    columns=flexible,
    basicstyle={\scriptsize\ttfamily},
    numberstyle=\emptyaccsupp,
    keywordstyle=\color{blue},
    commentstyle=\color{dkgreen},
    stringstyle=\color{mauve},
    frame=single,
    breaklines=true,
    breakatwhitespace=true,
    tabsize=4,
}

\lstdefinestyle{xsmall_shellblock}{
    backgroundcolor=\color{lightbg},
    language=bash,
    aboveskip=1mm,
    belowskip=1mm,
    showstringspaces=false,
    columns=flexible,
    basicstyle={\tiny\ttfamily},
    numberstyle=\emptyaccsupp,
    keywordstyle=\color{blue},
    commentstyle=\color{dkgreen},
    stringstyle=\color{mauve},
    frame=single,
    breaklines=true,
    breakatwhitespace=true,
    tabsize=4,
}

\lstnewenvironment{shell_snippet}[1][]
{%
    \lstset{style=shellblock, #1} %
}
{}

\lstnewenvironment{snippet}[1][]
{%
    \lstset{style=cppblock, #1} %
}
{}

\lstnewenvironment{xsmall_snippet}[1][]
{%
    \lstset{style=xsmall_cppblock, #1} %
}
{}

\lstnewenvironment{nsnippet}[1][]
{ %
    \lstset{style=block, numbers=left, #1}%
}
{}

\setlist{nosep} 

\begin{document}

    \title{Preliminary report:\linebreak{}Initial evaluation of StdPar implementations on AMD GPUs for HPC}
    \author[1]{Wei-Chen Lin, Simon McIntosh-Smith, Tom Deakin}
    \affil[1]{Department of Computer Science \protect\\ University of Bristol \protect\\ Bristol, UK \protect\\ wl14928@bristol.ac.uk, tom.deakin@bristol.ac.uk}
    \maketitle

    \section{Introduction}\label{sec:intro}

    Recently, AMD platforms have not supported offloading C++17 PSTL (\textit{StdPar}) programs to the GPU\@.
    Our previous work\cite{pmbs22-cpp} highlights how StdPar is able to achieve good performance across NVIDIA and Intel GPU platforms.
    In that work, we acknowledged AMD's past effort such as HCC, which unfortunately is deprecated and does not support newer hardware platforms.

    Recent developments by AMD, Codeplay, and AdaptiveCpp (previously known as hipSYCL or OpenSYCL) have enabled multiple paths for StdPar programs to run on AMD GPUs.
    This informal report discusses our experiences and evaluation of currently available StdPar implementations for AMD GPUs.
    We conduct benchmarks using our suite of HPC mini-apps with ports in many heterogeneous programming models, including StdPar.
    We then compare the performance of StdPar, using all available StdPar compilers, to contemporary heterogeneous programming models supported on AMD GPUs: HIP, OpenCL, Thrust, Kokkos, OpenMP, SYCL\@.
    Where appropriate, we discuss issues encountered and workarounds applied during our evaluation.

    Finally, the StdPar model discussed in this report largely depends on Unified Shared Memory (USM) performance and very few AMD GPUs have proper support for this feature.
    As such, this report demonstrates a proof-of-concept host-side userspace pagefault solution for models that use the HIP API\@.
    We discuss performance improvements achieved with our solution using the same set of benchmarks.

    \section{StdPar implementations}\label{sec:stdpar-implementations}
    The C++ semantics of StdPar programs dictate that all access to memory is inside a single address space.
For StdPar to work on discrete accelerators with a separate pool of memory, the vendor driver must support some form of unified memory (e.g.\ UVM in CUDA, XNACK in HSA) to adhere to C++ memory model.
As such, the performance of StdPar is directly tied to how well the unified memory implementation handles page migration and data residency.

This section introduces our StdPar implementations for AMD GPUs.
We discuss how each implementation handles memory access between the host and the device.

\subsection{AdaptiveCpp (hipSYCL) StdPar}\label{subsec:hipsycl-stdpar}

The AdaptiveCpp (recently renamed from hipSYCL) project is an independent SYCL implementation with support of the following platforms: CUDA, HIP, OpenCL (SPIR-V ingestion required), LevelZero, and OpenMP\@.
The project recently gained experimental support for StdPar\footnote{\url{https://github.com/OpenSYCL/OpenSYCL/blob/12f8c24d27c2e33e7357bb1bc44a2d12e60f427b/doc/stdpar.md}} by reusing major parts of the existing SYCL offloading infrastructure.
The changes required for StdPar are mostly contained in the compiler frontend; AdaptiveCpp enables StdPar on the same set of platforms it supports with SYCL, which includes the HIP backend\@.

To support StdPar's address space requirement, AdaptiveCpp replaces all memory-related functions in the program with HIP's unified memory API (\texttt{hipMallocManaged}) calls.
This essentially forces all allocations on the host to be controlled by AMD's KFD kernel driver.
A compiler flag (\texttt{--opensycl-stdpar-system-usm}) can be used to disable this replacement behaviour on systems with either 1) coherent host and device memory access or 2) a kernel with HMM support.

\subsection{ROCm StdPar}\label{subsec:roc-stdpar}

ROCm StdPar\footnote{\url{https://github.com/ROCmSoftwarePlatform/roc-stdpar}} (rocStdPar hereafter) is a new experimental StdPar implementation from AMD\@.
The implementation currently uses LLVM's HIP support and delegates most algorithms to ROCm's rocThrust implementation.
The compiler can be built without any ROCm components; it currently exists as a patch to upstream LLVM and a single-file glue header to rocThrust.

Like AdaptiveCpp, ROCm StdPar satisfies the single address space requirement by either substituting memory-related functions with HIP ones (enabled with \texttt{--hipstdpar-interpose-alloc}) or expects host-device memory to be coherent.

\subsection{Intel DPC++ w/ vendor plugin}\label{subsec:intel-dpcpp}

Intel's DPC++ (ICPX) is a fork of LLVM that adds support for the SYCL programming model.
The DPC++ runtime is designed in a manner that allows the development of backend plugins, enabling the execution of SYCL programs on various vendor platforms.
Presently, Intel's subsidiary, Codeplay, maintains CUDA and experimental HIP backends for DPC++.
Using Codeplay's vendor plugin, we can run SYCL programs on AMD GPUs\footnote{\url{https://developer.codeplay.com/products/oneapi/amd/2023.2.1/guides/}}.

For StdPar, our past study has shown that Intel's oneDPL header-only library is able to bridge the gap by implementing StdPar on top of SYCL\cite{pmbs22-cpp}.
For oneDPL to work while keeping the program ISO C++ compliant, we use a small shim header that replaces memory allocation functions with SYCL's USM allocations (i.e. \texttt{sycl::malloc\_shared}).
SYCL's USM allocation is backed by \texttt{hipMallocManaged} from the HIP API\@.
Like AdaptiveCpp, allocations implemented this way allow the host and device to share the same address space.

    \section{Mini-apps in the evaluation}\label{sec:mini-apps-in-the-evaluation}
    \begin{table}[ht]
    \caption{Mini-app benchmark configurations}
    \begin{adjustbox}{width=1\linewidth}
        \begin{tabular}{@{}lllll@{}}
            \toprule
            Mini-app & Input deck & Grid size & Steps & \begin{tabular}[c]{@{}l@{}}
                                                            Total Memory\\ Requirement
            \end{tabular} \\
            \midrule
            \multirow{1}{*}{BabelStream}
            & N/A    & N/A  & 100 & 12.9 GB          \\
            \midrule
            \multirow{1}{*}{miniBUDE}
            & bm1    & N/A  & 8   & 271 KB (n=65536) \\
            \midrule
            \multirow{1}{*}{TeaLeaf}
            & BM5@4k & 4000 & 2   & 1.96 GB          \\
            \midrule
            \multirow{1}{*}{CloverLeaf}
            & BM16   & 3840 & 300 & 2.95 GB          \\
            \bottomrule
        \end{tabular}
    \end{adjustbox}
    \label{tab:app-input-decks}
\end{table}

For a comprehensive evaluation, we select a range of HPC mini-apps that cover two main scenarios: compute bound and memory-bandwidth bound applications.
This section introduces the mini-apps used in our evaluation.

\subsection{BabelStream}\label{subsec:app-babelstream}

BabelStream\footnote{\url{https://github.com/UoB-HPC/BabelStream/tree/option_for_vec}} implements the standard McCalpin STREAM benchmark with an additional Dot product kernel in a wide range of programming models~\cite{mccalpin:1995,babelstream:2018}.
This memory-bandwidth bound benchmark measures the time taken for each unique kernel and generates memory bandwidth data in MB/s.

As shown in \cref{tab:app-input-decks}, we use the default iteration count of 100 and set the array size to $ 2^{29} $($\approx$ 4GB) to avoid any unrealistic caching behaviours.

\subsection{miniBUDE}\label{subsec:app-minibude}

MiniBUDE \footnote{\url{https://github.com/UoB-HPC/miniBUDE/tree/v2}} is a molecular docking benchmark that is reduced from the full scale Bristol University Docking Enging (BUDE).
The mini-app implements the virtual screening process where we evaluate energy values from docking ligand and protein molecules in different poses.
This application is compute-bound, primarily because the input decks typically occupy very little memory (i.e. $approx$ 300KB).
Structured similarly to BabelStream, miniBUDE is also implemented in a wide range of programming models.

MiniBUDE exposes two tuning variables: \texttt{wgsize} and \texttt{PPWI}.
Variable \texttt{wgsize} controls the hierarchical kernel launch's group size (e.g.\ the workgroup size of an NDRange launch in OpenCL terminology).
StdPar and Thrust does not yet expose a way to express hierarchical parallelism, so \texttt{wgsize} is not tunable for these two models.
The \texttt{PPWI} variable controls the number of poses per task (e.g. workitem in OpenCL terminology).
\texttt{PPWI} is used in the main kernel's innermost loop, which is statically unrolled using C++ templates; this variable is supported in all models.

For our evaluation, we use the BM1 input, with iteration count shown in \cref{tab:app-input-decks}.
BM1 is a small problem size with just 938 proteins and 26 ligands.

\subsection{CloverLeaf}\label{subsec:app-cloverleaf}

CloverLeaf is a hydrodynamics mini-app that solves the compressible Euler Equations using a structured grid.
This is a complex mini-app with more than 100 unique kernels.
The fluid simulation is done in configurable resolution and timestamps.
Each timestep involves two main categories of computation: 2D reductions and 2D grid traversal.
Like BabelStream, CloverLeaf has been ported to multiple programming models.

CloverLeaf's kernel submission frequency is high, and each kernel's data dependencies are complex: many involve more than a dozen independent buffers.

\subsection{TeaLeaf}\label{subsec:app-tealeaf}

TeaLeaf is a diffusion mini-app that solves the heat conduction equations.
Like CloverLeaf, this is a complex mini-app with a high number of unique kernels.
The principal access pattern uses a 5-point stencil to compute the Sparse Matrix Vector Product (SpMV) of our grid.

For our evaluation, we use the Conjugate Gradient solver.
This solver, in addition to SpMV, requires multiple 2D reductions at each step which can help identify any performance weaknesses of the underlying programming model or implementation.

    \section{USM (XNACK) on AMD GPUs}\label{sec:usm-on-amd-gpus}
    \begin{table*}[ht!]
    \centering
    \caption{Platform details}
    \begin{adjustbox}{width=1\linewidth}
    \begin{tabular}{ lllllll}
        \toprule
         Name & Architecture & Abbreviation & Attachment & XNACK
        & \begin{tabular}[c]{@{}l@{}} Theoretical Peak \\Mem. Bandwidth \\(GB/s) \end{tabular}
        &\begin{tabular}[c]{@{}l@{}} Theoretical Peak \\FP32 FLOP/s\\(GFLOP/s)\end{tabular}           \\
        \hline
        AMD Instinct MI100           & CDNA (gfx908)         & MI100       & PCIe 4.0 x16  & Disabled (kernel)    & 1228   & 23070  \\
        AMD Radeon VII               & Vega 20 (gfx906)      & RadeonVII   & PCIe 3.0 x16  & Enabled (kernel)     & 1024   & 13800  \\
        AMD Ryzen 7950X IGP/APU      & Navi 2 (gfx1036)      & Raphael     & PCIe 4.0 x8   & Unsupported          & 96     & 563.2  \\
        \bottomrule
    \end{tabular}
    \end{adjustbox}
    \label{tab:platforms}
\end{table*}

    To be able to use a host pointer in a device kernel, the GPU hardware must be able to signal a pagefault to the host and retry the access once the page has been migrated to GPU memory in some way.
On AMD GPUs, this hardware feature is called XNACK\@.
Currently, only HPC and a handful of consumer GPUs support this feature.

Enabling XNACK on AMD GPUs requires a somewhat recent Linux kernel along with a non-default kernel argument: \texttt{amdgpu.noretry=0}.
As of Linux kernel 6.2 (mainline), XNACK still appears to be a moving target in terms of stability; earlier (5.15) kernels exhibit random panics and hangs that require a physical power-cycle to recover.
Documentation on XNACK is virtually non-existent beyond kernel source code and third-party discussions\footnote{https://niconiconi.neocities.org/tech-notes/xnack-on-amd-gpus/}.

Due to the factors discussed above, very few production clusters that have AMD GPUs are configured with XNACK enabled.
Currently, only ORNL's Frontier and the Crusher testbed appear to have any support for XNACK\@.
We are uncertain if LUMI has this feature enabled.
Requesting access to these machines has a lead-time of up to 6 months.

\subsection{USM without XNACK}\label{subsec:utpx}

Without XNACK, AMD GPU devices cannot handle pagefaults and ROCm degrades the allocation to host-resident memory where all accesses from the device must cross the host-device interconnect (e.g., PCIe).
In this degraded mode, no page migration occurs, so performance for memory-bandwidth bound applications will be limited to interconnect performance.
For example, a MI100 operating at PCIe 4.0 x16 without XNACK will see application performance capped at 31.5GB/s when device-resident memory should be capable of 1228.8 GB/s, a 40x difference.

To overcome this, we implement a simple \texttt{LD\_PRELOAD} program called UTPX (Userspace Transparent Paging Extension)\footnote{\url{https://github.com/UoB-HPC/utpx}}.
UTPX is a proof-of-concept shim program that accelerates HIP managed allocations (e.g. \texttt{hipMallocManaged}) on systems without XNACK or with XNACK disabled.

\subsection{UTPX design}\label{subsec:utpx-design}

\begin{figure}[ht!]
    \centerline{\includegraphics[width=1\linewidth]{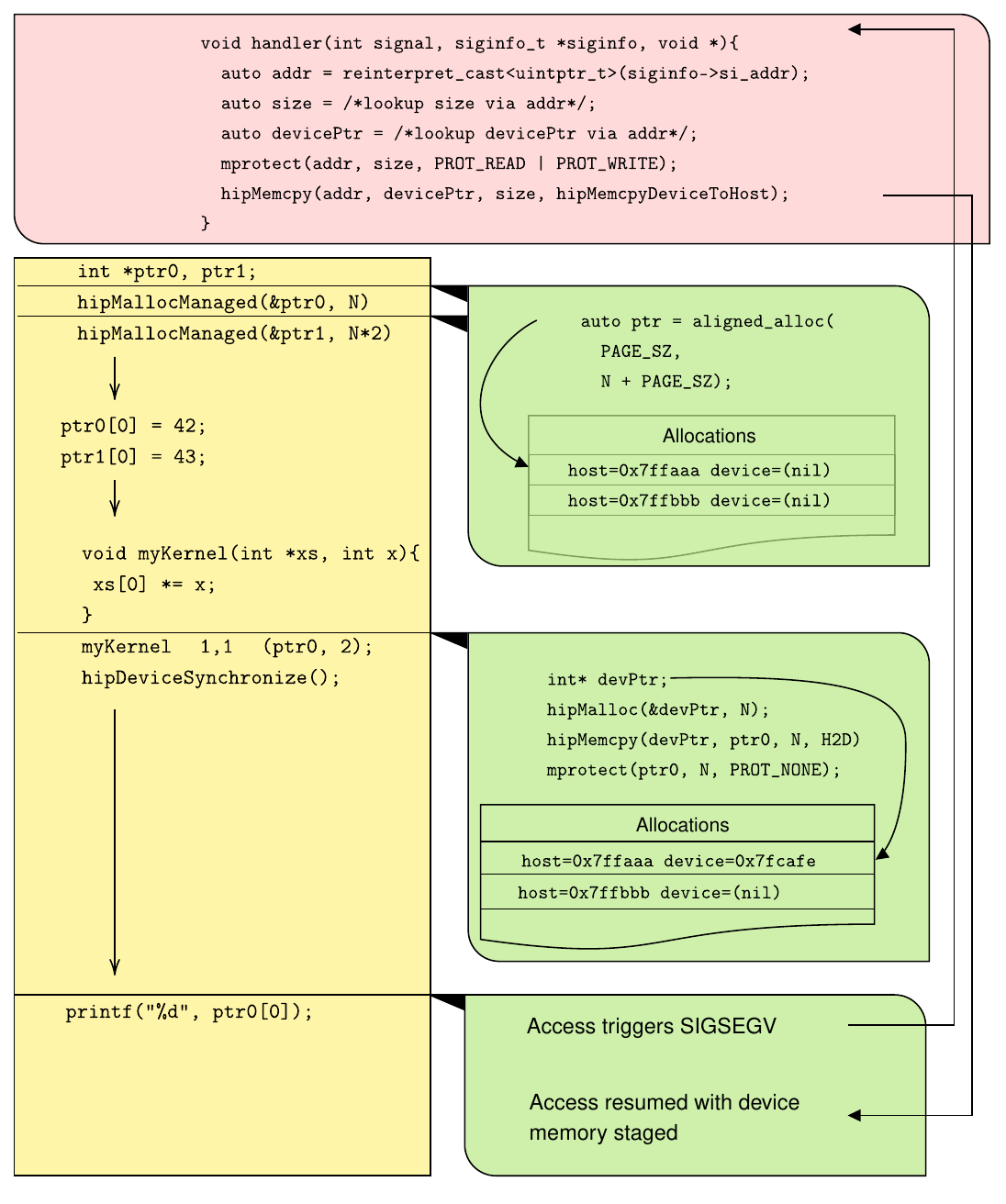}}
    \caption{UTPX HIP API interposing sequence}
    \label{fig:utpx-diagram}
\end{figure}

UTPX solves USM by shifting pagefault handling to the host where it can be easily accomplished in userspace.
The implementation performs userspace page migration using \texttt{mprotect} and signal handlers.
UTPX is an \texttt{LD\_PRELOAD} program, and as such, can be used on existing programs without any recompilation.

Paging is done at the granularity of the allocation itself using a Mirror-on-Access (mirror) scheme.
With mirror, initial allocations are resident on the host, and a separate device-resident allocation is made whenever a kernel is launched that has a dependency on the allocation.
A device to host write-back is triggered by an \texttt{mprotect} induced pagefault, this happens if the host copy of the device-resident memory is accessed in any way (e.g., through pointer dereference).

As an alternative to the mirror scheme mentioned above, we also implement two alternative schemes of memory management:
\begin{itemize}
    \item The \textit{device} scheme replaces all allocations with \texttt{hipMalloc}, effectively making all memory allocations device resident.
    The userspace pagefault handler is not installed here because allocations made with \textit{hipMalloc} are host-visible.
    \item The \textit{advise} scheme does not replace \texttt{hipMallocManaged}, but inserts extra \texttt{hipMemAdvise} on allocation calls and \texttt{hipMemPrefetchAsync} to kernel launches;
    on each kernel launch, instead of creating device allocations, we simply prefetch memory from the host to the device.
\end{itemize}

Per-kernel memory dependency is resolved by introspection of the \texttt{.note} section of the HSA code object (HSACO) ELF image.
This ELF image is available if we carefully intercept \texttt{hsa\_code\_object\_reader\_create\_from\_memory} from \texttt{LD\_PRELOAD}.
For lambda capture objects, we perform a ranged scan of pointer values at a 2-byte increment against all known past allocations.
A pointer scan is required because the program binary does not store layout information for structures.

It's worth noting that our current implementation does not support pointer indirection beyond the first level.
This limitation is relevant for complex data structures like \texttt{std::vector}.
However, a straightforward approach of chasing pointers (such as treating each argument as roots in a garbage collector) could be employed to implement this.

To associate a kernel launch-site (e.g. $\lll\ggg$/\texttt{hipLaunchKernelGGL}) with the correct set of metadata, we intercept \texttt{\_\_hipRegisterFunction} which includes the pointer to the kernel function and the mangled kernel name.
If the program loads the kernel image at runtime, as implemented in ICPX's HIP plugin, we make use of the fact that a \texttt{hipFunction\_t} is internally a pointer to the \texttt{amd::DeviceFunc} structure.
This structure stores the mangled name of the kernel as a \texttt{std::string} at a 144 byte offset from the base of the pointer.
Once the mangled kernel name is obtained, we can find the correct argument information by doing a lookup with metadata obtained from the ELF image.

    \section{Performance results}\label{sec:performance-results}
    \begin{table}[ht!]
    \caption{Software versions and configuration, MI100* is a separate HPE prototype system}
    \small
    \begin{adjustbox}{width=1\linewidth}
        \begin{tabular}{rl}
            \hline
            Software       & Version           \\ \hline
            OS & \begin{tabular}[c]{@{}l@{}}
                     RadeonVII: Ubuntu 22.04 LTS\\ MI100: RHEL 8.6\\ MI100*: SLES 15 SP5 \\ Raphael: Fedora 37
            \end{tabular} \\ \cline{2-2}
            Kernel & \begin{tabular}[c]{@{}l@{}}
                         RadeonVII: 6.4.6+HMM, \texttt{amdgpu.noretry=0}\\ MI100: 4.18+HMM\\ MI100*: 5.14+HMM, \texttt{amdgpu.noretry=0}\\ Raphael: 6.4.15+HMM
            \end{tabular} \\ \cline{2-2}
            \begin{tabular}[c]{@{}r@{}}
                ROCm\\ (w/ rocThrust\\ rocPRIM)
            \end{tabular} & \begin{tabular}[c]{@{}l@{}}
                                RadeonVII: 5.5.1 (LLVM 16)\\ MI100: 5.4.1 (LLVM15) \\ MI100*: 5.5.1 (LLVM 16)\\ Raphael: 5.5.1 (LLVM 16)
            \end{tabular} \\ \cline{2-2}
            GCC, libstdc++ & 12.3.0 \\ \cline{2-2}
            AOMP & 18.0.0 (LLVM 18) \\ \cline{2-2}
            ICPX, oneDPL & 2023.2.1  (LLVM 17) \\ \cline{2-2}
            AdaptiveCpp & \begin{tabular}[c]{@{}l@{}}
                          SHA: fd5d1c\\ LLVM 18, SHA: ecb855a5
            \end{tabular} \\ \cline{2-2}
            ROCm StdPar & \begin{tabular}[c]{@{}l@{}}
                              SHA: e900d4\\ LLVM 18, SHA: ecb855a5
            \end{tabular} \\ \hline
        \end{tabular}
    \end{adjustbox}
    \label{tab:software-vers}
\end{table}

As this report is intended to provide early feedback for emerging StdPar models on AMD GPUs, we have opted to conduct benchmarks on a local RadeonVII test system.
RadeonVII is selected for its resemblance to MI50 and the official ROCm support from AMD\@.

For XNACK-disabled scenarios, we present single-node MI100 results obtained by the GW4 Isambard system.
We encountered difficulties in accessing AMD HPC GPUs with XNACK enabled.
However, we were able to locate a MI100 system with XNACK enabled in HPE's Grenoble prototype system.
Access to this system is only available for a limited time window, so we ran exclusively XNACK-enabled benchmarks to achieve XNACK-enabled coverage on MI100.

Specific details of our hardware selection are shown in \cref{tab:platforms}. and \cref{tab:software-vers} lists specific versions of the compilers and software used for benchmarks presented in this section.

The rocStdPar compiler supports two modes of operation.
We use \texttt{rocStdPar@ITP} for benchmarks that were compiled with \texttt{--hipstdpar-interpose-alloc} and \texttt{rocStdPar@HMM} for ones without.
The \texttt{--hipstdpar-interpose-alloc} flag replaces all memory allocations (e.g. \texttt{malloc}) with \texttt{hipMallocManaged} so that the pagefault behaviour does not require kernel-level HMM support.
In theory, AdaptiveCpp also supports this via the \texttt{-opensycl-stdpar-system-usm} but due to time constraints, we did not test this.

\subsection{Results: BabelStream}\label{subsec:results-babelstream}

\begin{figure}[ht!]
    \centerline{\includegraphics[width=1\linewidth]{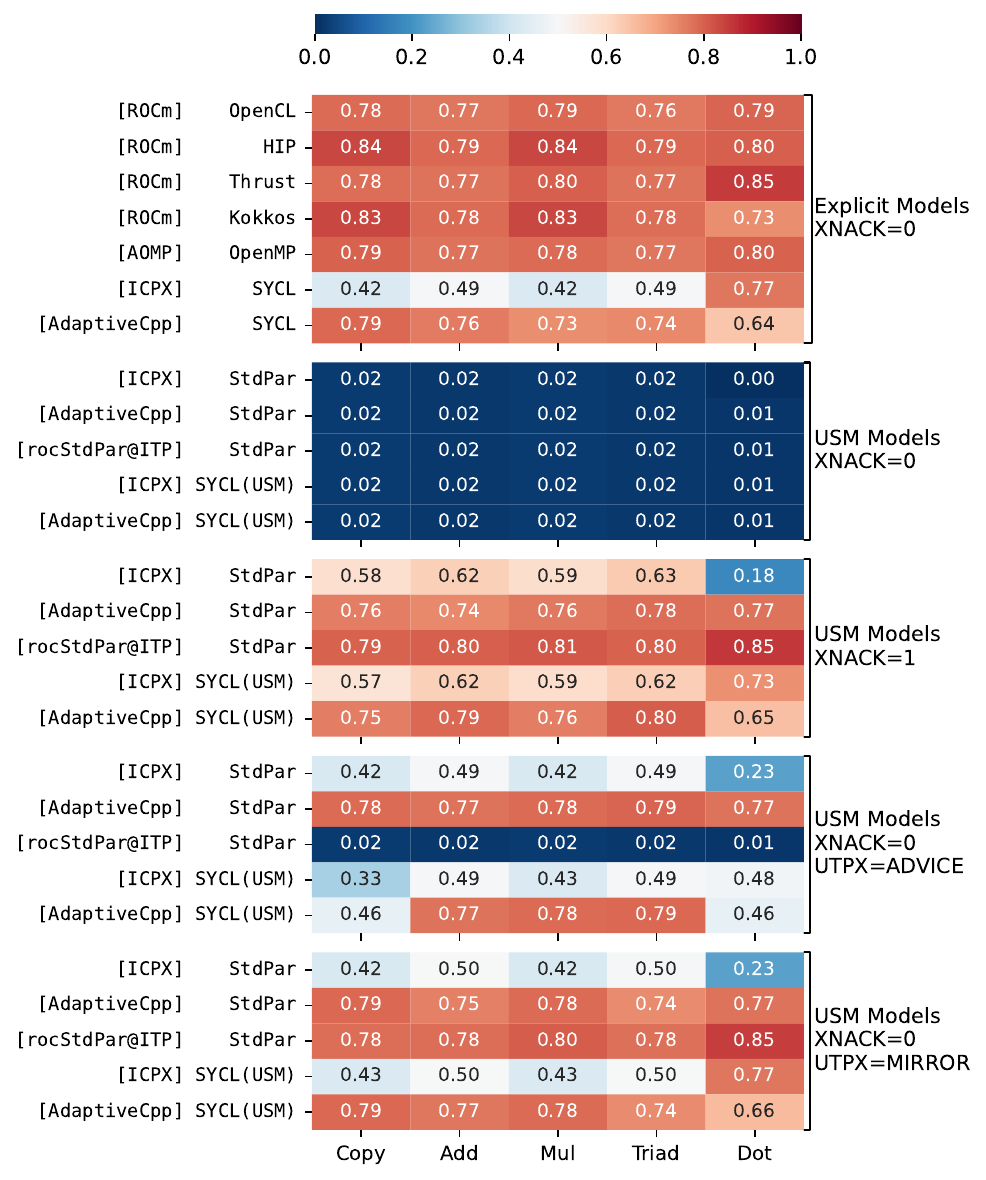}}
    \caption{BabelStream kernel bandwidth as fraction of theoretical peak on RadeonVII, \textbf{higher is better}}
    \label{fig:effective-babelstream-radeonvii}
\end{figure}

\begin{figure}[ht!]
    \centerline{\includegraphics[width=1\linewidth]{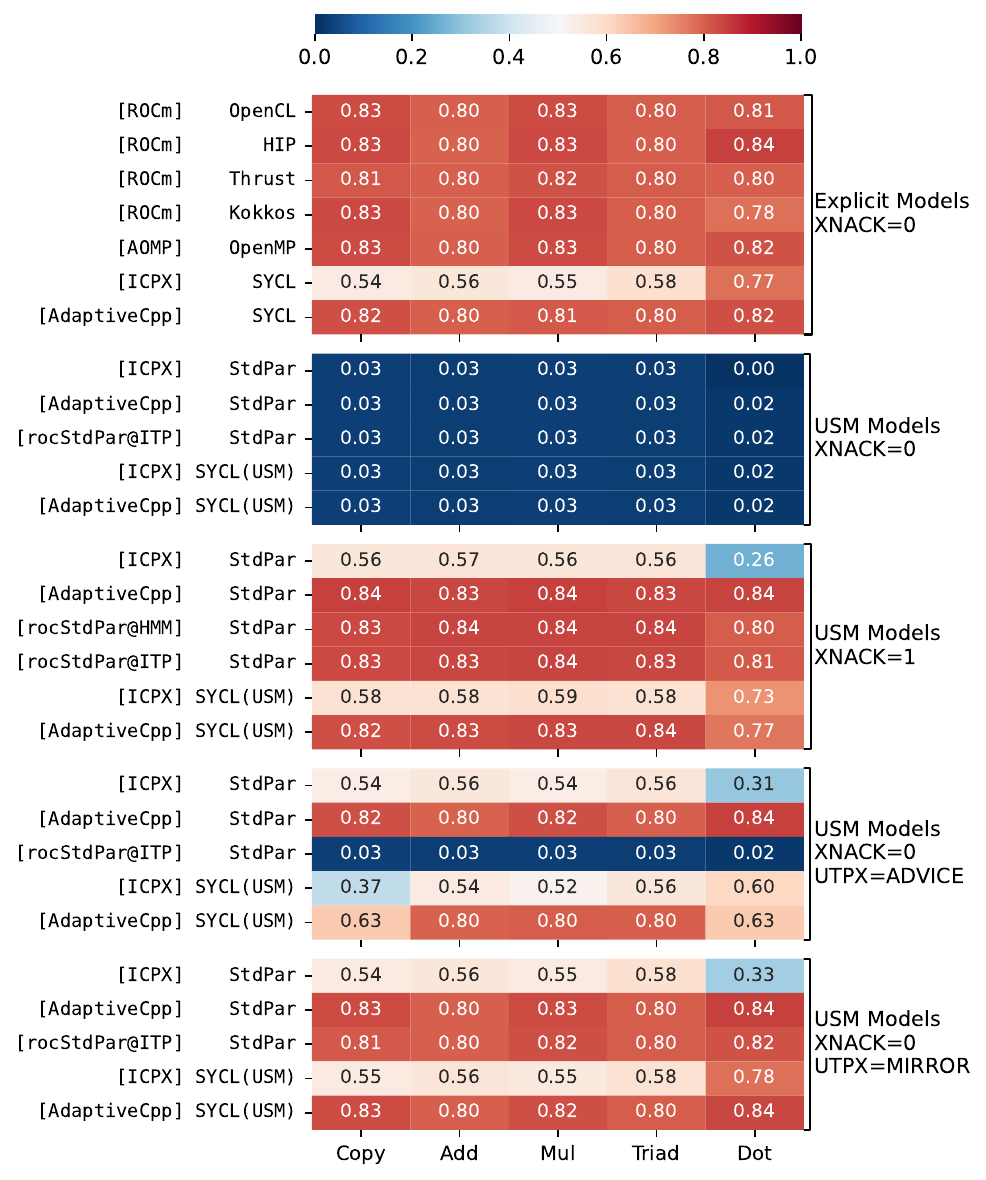}}
    \caption{BabelStream kernel bandwidth as fraction of theoretical peak on MI100, \textbf{higher is better}}
    \label{fig:effective-babelstream-mi100}
\end{figure}

\begin{figure}[ht!]
    \centerline{\includegraphics[width=1\linewidth]{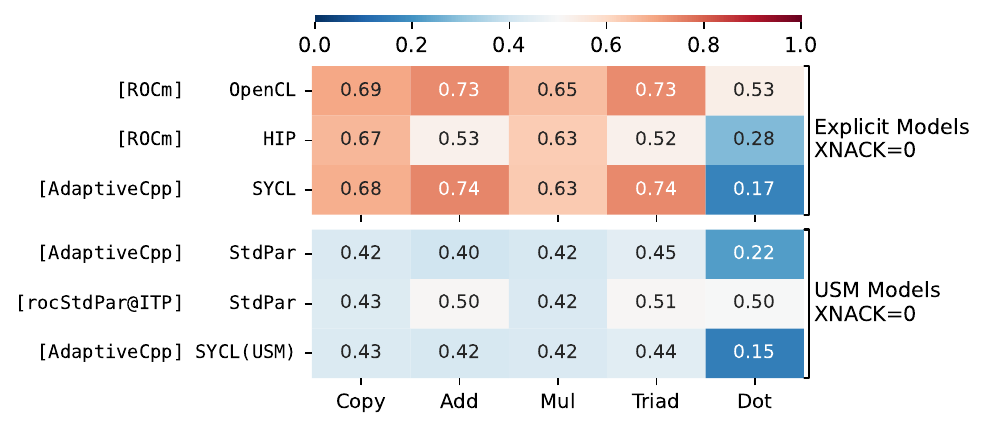}}
    \caption{BabelStream kernel bandwidth as fraction of theoretical peak on Raphael, \textbf{higher is better}}
    \label{fig:effective-babelstream-gfx1036}
\end{figure}

BabelStream results are mostly inline with expectations for a GPU platform: most explicit models are reaching about 80\% of the bandwidth on all five kernels.
USM models without XNACK, as shown in the second group from the top of \cref{fig:effective-babelstream-radeonvii} and \cref{fig:effective-babelstream-mi100}, are limited to PCIe bandwidth and match the documented behaviour of host-resident memory.
Intel's ICPX compiler as a whole appears suboptimal with Codeplay's HIP backend for memory-bandwidth bound kernels.

On RadeonVII, the use of XNACK does not appear to impose significant overhead: StdPar models on all three implementations performed nearly on-par with other more established models.
Likewise, on both RadeonVII and MI100, UTPX in mirror mode successfully restored near full performance for all USM models.
The effectiveness of UTPX is unexpected; we expected that userspace pagefault handling would have significant overhead, and adding an extra layer of indirection for all kernel launches would also add severe latency to the program.

Results for rocStdPar presented in this section fail validation.
Looking at the memory allocation logic of rocStdPar, it employs a fairly complex allocation scheme when the interposing mode is enabled:
rocStdPar attempts to allocate a large page-aligned block of memory for both bookkeeping and as the primary allocation on top of \texttt{hipMallocManaged}.
When UTPX is used together with rocStdPar on large allocations (> 1GB), the host write-back (see \cref{fig:utpx-diagram}) on pagefault seems to copy pages that are corrupted.
It's unclear why this only occurs for large allocations.

UTPX's advice mode failed to provide a meaningful uplift for rocStdPar, with performance equal to the bandwidth of XNACK and UTPX disabled results.
We suspect the complex allocation scheme broke ROCm's undocumented alignment invariants required for \texttt{hipMemPrefetchAsync} and \texttt{hipMemAdvise} to function correctly.
A simple validation was done where we replace the allocation behaviour of rocStdPar with just \texttt{hipMallocManaged} and nothing else.
With this change, advice mode was able to match the performance of mirror mode, and BabelStream passes validation.

Device mode was tested with BabelStream reporting bandwidth figures on-par with explicit models.
However, it's important to note that BabelStream does not time the final device-to-host transfer, which is needed for validation.
Consequently, while the bandwidth measured during kernel execution accurately reflects the performance, device mode introduces additional data movement time that can dominate the overall benchmark results.
In light of this, we have omitted device mode results as we are unsure how this would fit with existing bandwidth measurements.
An issue\footnote{\url{https://github.com/UoB-HPC/BabelStream/issues/161}} has been opened in the BabelStream repository for us to assess how the transfer timing information can be included in the output of the benchmark.

Results on the Raphael APUs, as shown in \cref{fig:effective-babelstream-gfx1036}, stand out.
Explicit models show a significant lead over USM ones, despite the GPU sharing the memory with the host.
After consulting with AMD and reviewing the relevant hardware topology documentations, we believe the main cause of this was due to the limited attachment bandwidth;
the GPU communicates with the CPU using an on-die PCIe 4.0 x8 connection.

\begin{figure}[h!]
    \centerline{\includegraphics[width=1\linewidth]{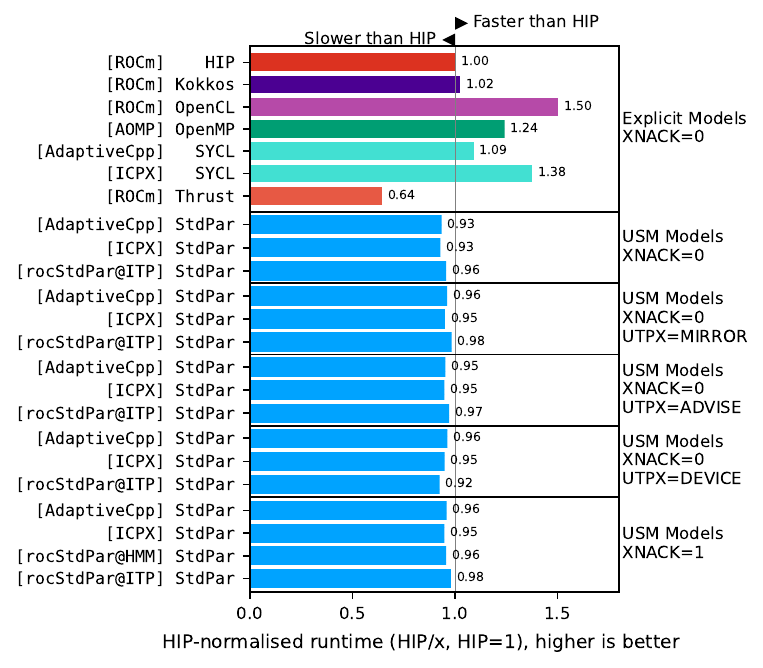}}
    \caption{miniBUDE normalised runtime across all models against HIP on RadeonVII, \textbf{lower is better}}
    \label{fig:effective-minibude-radeonvii}
\end{figure}

\begin{figure}[h!]
    \centerline{\includegraphics[width=1\linewidth]{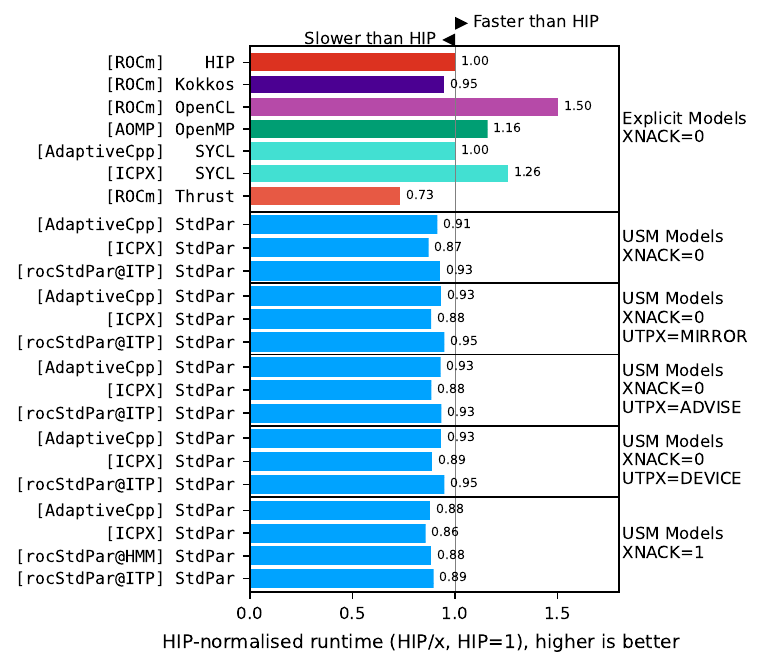}}
    \caption{miniBUDE normalised runtime across all models against HIP on MI100, \textbf{lower is better}}
    \label{fig:effective-minibude-mi100}
\end{figure}

\begin{figure}[h!]
    \centerline{\includegraphics[width=1\linewidth]{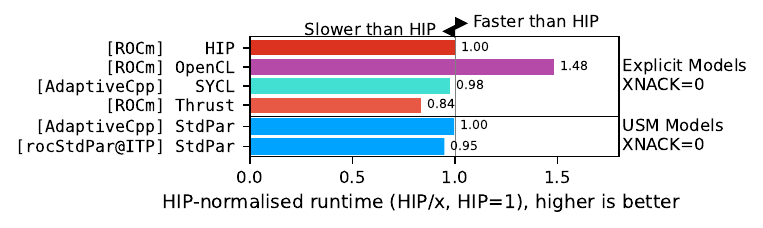}}
    \caption{miniBUDE normalised runtime across all models against HIP on Raphael, \textbf{lower is better}}
    \label{fig:effective-minibude-gfx1036}
\end{figure}

\subsection{Results: miniBUDE}\label{subsec:results-minibude}

\begin{figure*}[htpb!]
    \centerline{\includegraphics[width=1\linewidth]{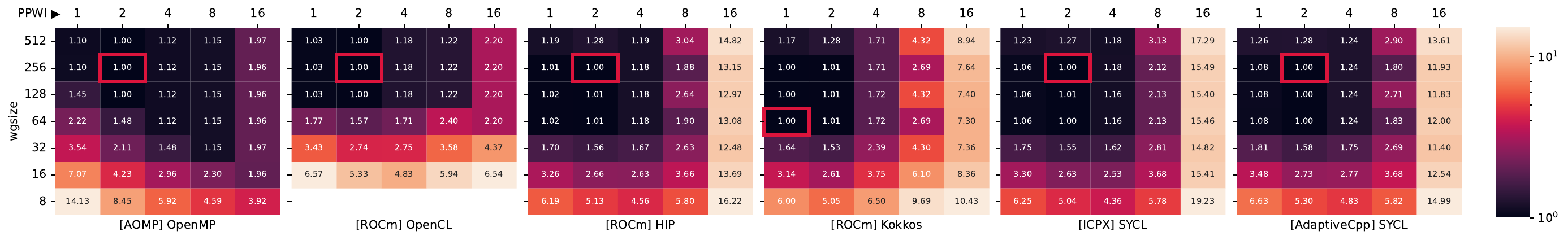}}
    \caption{miniBUDE PPWI/wgsize tuning on RadeonVII, fastest combination outlined in red, \textbf{lower is better}}
    \label{fig:tuning-minibude-grid-radeonvii}
\end{figure*}

\begin{figure*}[htpb!]
    \centerline{\includegraphics[width=1\linewidth]{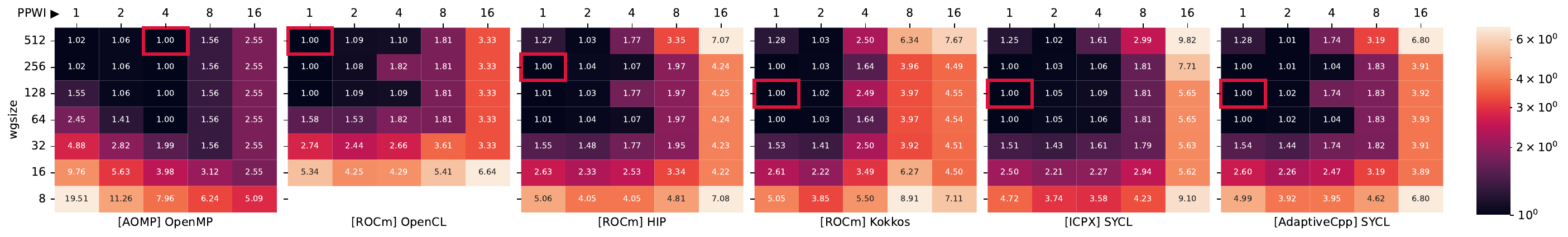}}
    \caption{miniBUDE PPWI/wgsize tuning on MI100, fastest combination outlined in red, \textbf{lower is better}}
    \label{fig:tuning-minibude-grid-mi100}
\end{figure*}

\begin{figure*}[htpb!]
    \centering
    \begin{minipage}{.5\linewidth}
        \centerline{\includegraphics[width=1\linewidth]{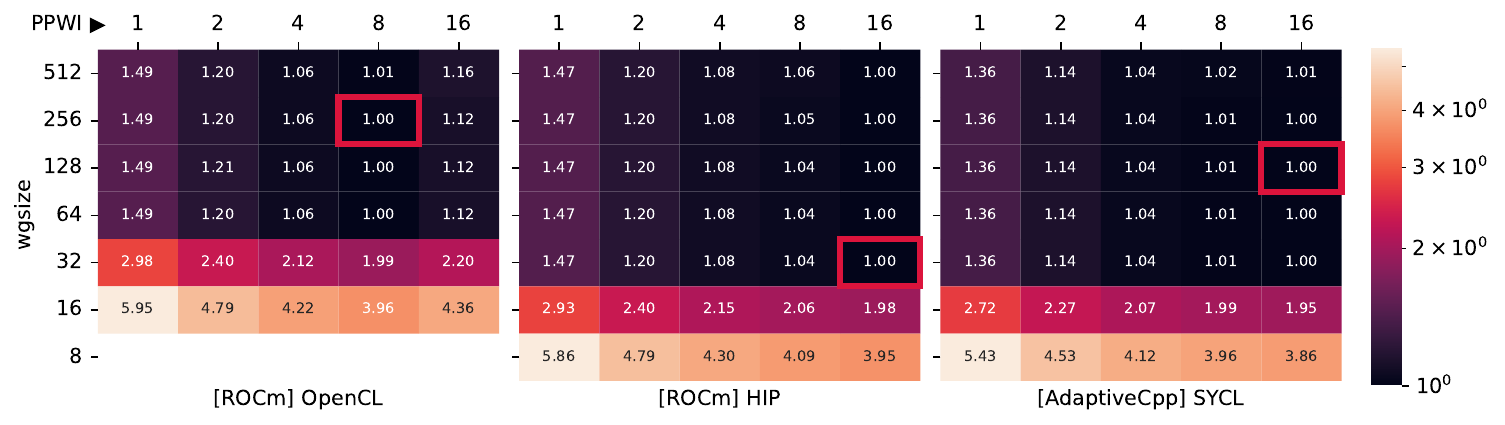}}
        \caption{miniBUDE PPWI/wgsize tuning on Raphael, fastest combination outlined in red, \textbf{lower is better}}
        \label{fig:tuning-minibude-grid-gfx1036}
    \end{minipage}
    \begin{minipage}{.48\linewidth}
        \begin{subfigure}{.33\textwidth}
            \centerline{\includegraphics[width=1\linewidth]{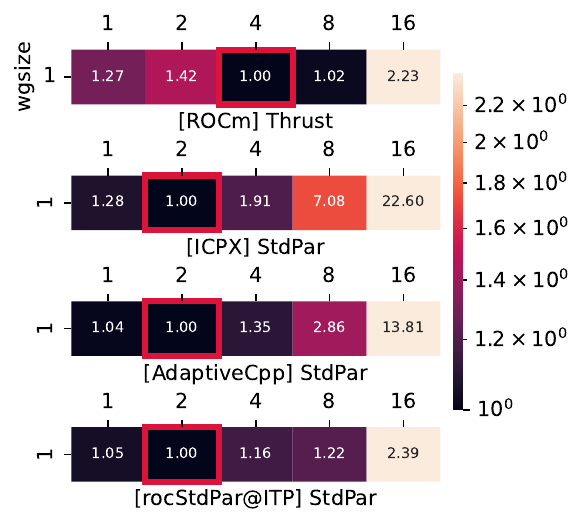}}
            \caption{RadeonVII}
            \label{fig:tuning-minibude-flat-radeonvii}
        \end{subfigure}
        \begin{subfigure}{.33\textwidth}
            \centerline{\includegraphics[width=1\linewidth]{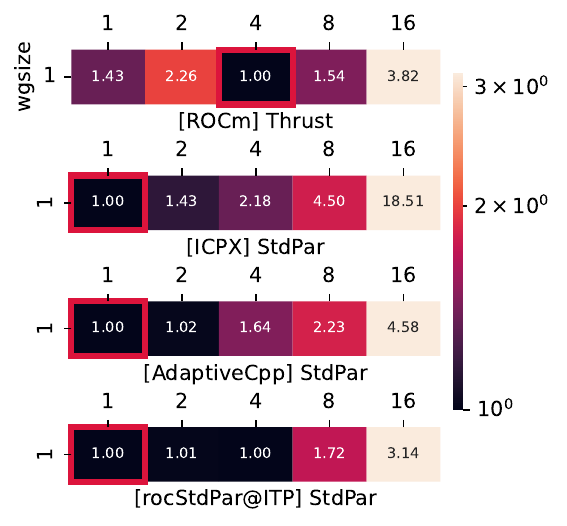}}
            \caption{MI100}
            \label{fig:tuning-minibude-flat-mi100}
        \end{subfigure}%
        \begin{subfigure}{.33\textwidth}
            \centerline{\includegraphics[width=1\linewidth]{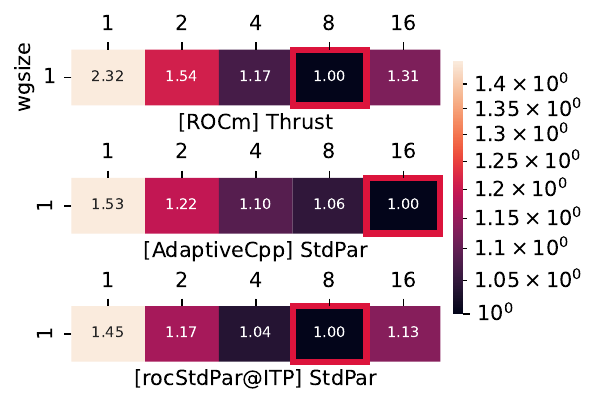}}
            \caption{Raphael}
            \label{fig:tuning-minibude-flat-gfx1036}
        \end{subfigure}
        \caption{miniBUDE PPWI/wgsize tuning for models without wgsize parameter, fastest combination outlined in red, \textbf{lower is better}}
    \end{minipage}
\end{figure*}

MiniBUDE results are presented in two parts to show 1) effects of the tuning parameters, or the lack thereof, on specific models, and 2) the effective performance after tuning.

\cref{fig:tuning-minibude-grid-radeonvii}, \cref{fig:tuning-minibude-grid-mi100}, \cref{fig:tuning-minibude-grid-gfx1036} present heatmaps where the X axis scales PPWI and Y axis scales the wgsize parameter.
This benchmark highlights a crucial tradeoff that models like StdPar and Thrust made: workgroup size is an implementation detail and not programmer adjustable.
Across all tuning results shown here, the best performing PPWI and wgsize combination changes for different platforms and programming models.
In certain cases, selecting parameters that are immediately adjacent in any direction to the best performing combination can see an up-to 50\% loss (\cref{fig:tuning-minibude-grid-mi100}).

Using the best performing PPWI and wgsize combination, we compare results from all models and compilers in \cref{fig:effective-minibude-radeonvii}, \cref{fig:effective-minibude-mi100}, and \cref{fig:effective-minibude-gfx1036}.
Here, the fastest model is OpenCL, even AMD's first party model, HIP, trailed behind.
This agrees with our past evaluations for miniBUDE\cite{julia:2021}.
We suspect the added complexity of C++-based abstractions is introducing unnecessary optimisation burdens to the main kernel.
MiniBUDE is highly sensitive to missed or poorly optimised code due to its loop structure;
the pose count will amplify any suboptimal code path in the ligand and atom inner loop.

StdPar models all performed about the same and matched the performance of HIP\@.
Surprisingly, Thrust (backed by rocThrust) was a lot worse compared with StdPar results.

While miniBUDE is not a memory-bandwidth bound benchmark, we still enable UTPX to gauge the relative overhead of intercepting HIP API calls to perform complex kernel argument manipulations.
The overall performance impact across both RadeonVII MI100 appears to be minimal.

Support for the Raphael APU is limited.
For ICPX, the compiler is missing an entry from the architecture table, and a bug has been opened to track this issue \footnote{https://github.com/intel/llvm/issues/11203}.
In general, we find downstream vendors are reluctant to support GPUs that are outside ROCm's support, we discuss why this is an issue in \cref{sec:rocm-experience}.
Possibly for the same reason, AMD's own AOMP compiler was unable to target the APU even though a PR to support this has already been merged \footnote{https://github.com/ROCm-Developer-Tools/aomp/pull/452}.
In this instance, AOMP appears to be missing the required math library.

\subsection{Results: CloverLeaf}\label{subsec:results-cloverleaf}

\begin{figure}[ht!]
    \centerline{\includegraphics[width=1\linewidth]{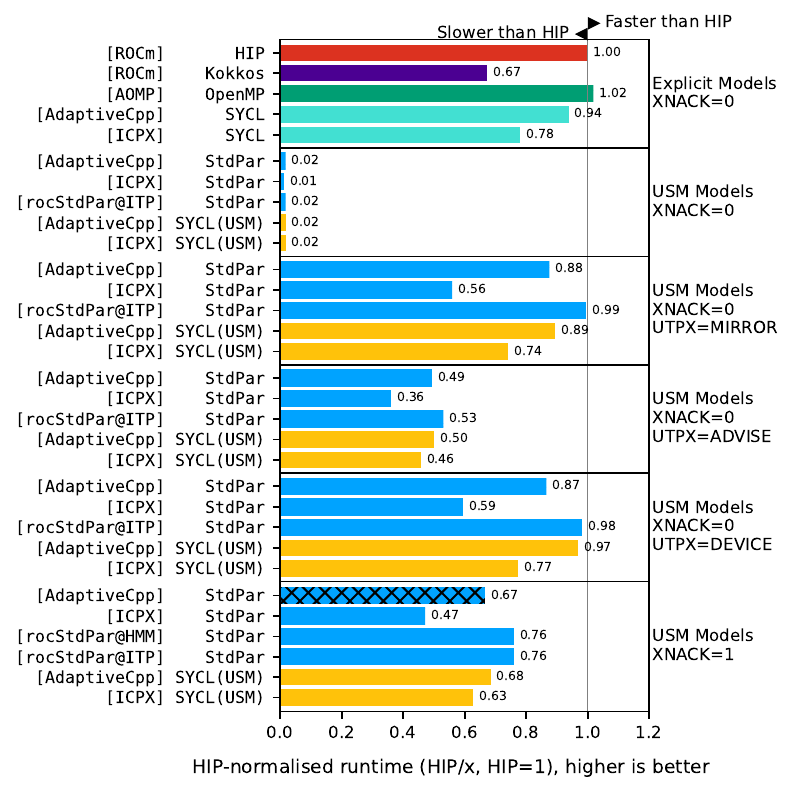}}
    \caption{CloverLeaf normalised runtime across all models against HIP on RadeonVII, \textbf{higher is better}}
    \label{fig:effective-cloverleaf-radeonvii}
\end{figure}

\begin{figure}[ht!]
    \centerline{\includegraphics[width=1\linewidth]{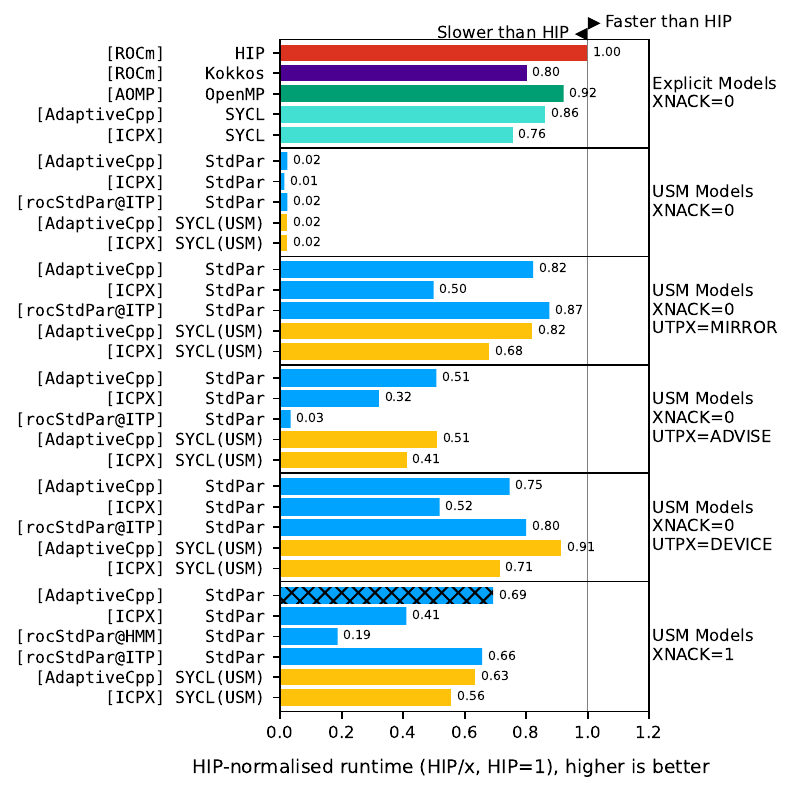}}
    \caption{CloverLeaf normalised runtime across all models against HIP on MI100, \textbf{higher is better}}
    \label{fig:effective-cloverleaf-mi100}
\end{figure}

\begin{figure}[ht!]
    \centerline{\includegraphics[width=1\linewidth]{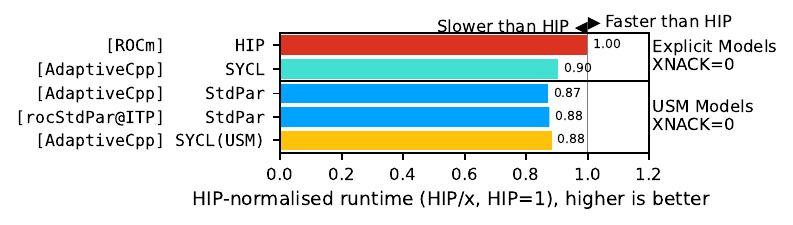}}
    \caption{CloverLeaf normalised runtime across all models against HIP on Raphael, \textbf{higher is better}}
    \label{fig:effective-cloverleaf-gfx1036}
\end{figure}

CloverLeaf results are presented in \cref{fig:effective-cloverleaf-gfx1036}, \cref{fig:effective-cloverleaf-radeonvii}, and \cref{fig:effective-cloverleaf-mi100}.
In these figures, all StdPar implementations were able to achieve approximately 70\% of HIP performance with XNACK enabled.
Performance without XNACK was notably poor, as expected due to the default host-resident behaviour of \texttt{hipMallocManaged}.
The lower performance of ICPX, as discussed in \cref{subsec:results-babelstream}, was also reflected here.

As with BabelStream, we applied our UTPX program for USM models, which resulted in a consistent and substantial performance improvement across with all UTPX modes.
This improvement not only brought performance closer to what was achieved with HIP but, in specific cases, even matched it.
Except ICPX, all StdPar models managed to outperform Kokkos and were roughly on par with the non-USM variant of SYCL\@.
Enabling interposing for rocStdPar did not alter the performance characteristic in any meaningful way.

For USM models of CloverLeaf, data access after the initial buffer setup and domain decomposition was almost entirely device-resident, with host access only required for reductions occurring every 20 time steps.
In such access patterns, the performance of UTPX in device mode was comparable to mirror mode.

Interestingly, both device and mirror mode outperformed the hardware-assisted XNACK and the less intrusive \textit{advise} mode.
For a device-resident application like CloverLeaf, the overhead of dynamic page fault management, whether done in software or hardware, may be challenging to recover without more advanced (preemptive or compiler-assisted) heuristics.
However, we found the performance of XNACK compared to software solutions like UTPX to be underwhelming.
This could potentially be attributed to the extra register usage, as the code was compiled without specific targeting of XNACK.
Additionally, it's worth noting that UTPX's advise mode consistently failed on different systems, as observed from the results on RadeonVII and MI100.
We suspect that ROCm failed to create an allocation with the correct alignment requirements at runtime.

On the Raphael APU, USM models, including all implementations of StdPar, achieved nearly 90\% of HIP performance.
This demonstrates that under ideal conditions with a shared address space between the CPU and the GPU, the model itself did not impose significant performance overheads.
This consistent overhead aligns with our previous studies on other APU platforms as discussed in our past work \cite{pmbs22-cpp}.
As discussed in \cref{subsec:results-minibude}, AOMP and ICPX do not yet support the Raphael APU\@.

Note that UTPX was developed in a relatively short (< 3 days) amount of time.
While benchmarking, we have identified an issue where the mirror mode would occasionally result in deadlocks on larger allocations.
Rerunning the benchmark usually succeeds immediately after a failed run.
We think the root cause is a stack value corruption that originates from within ROCm.

\subsection{Result: TeaLeaf}\label{subsec:results-tealeaf}

\begin{figure}[ht!]
    \centerline{\includegraphics[width=1\linewidth]{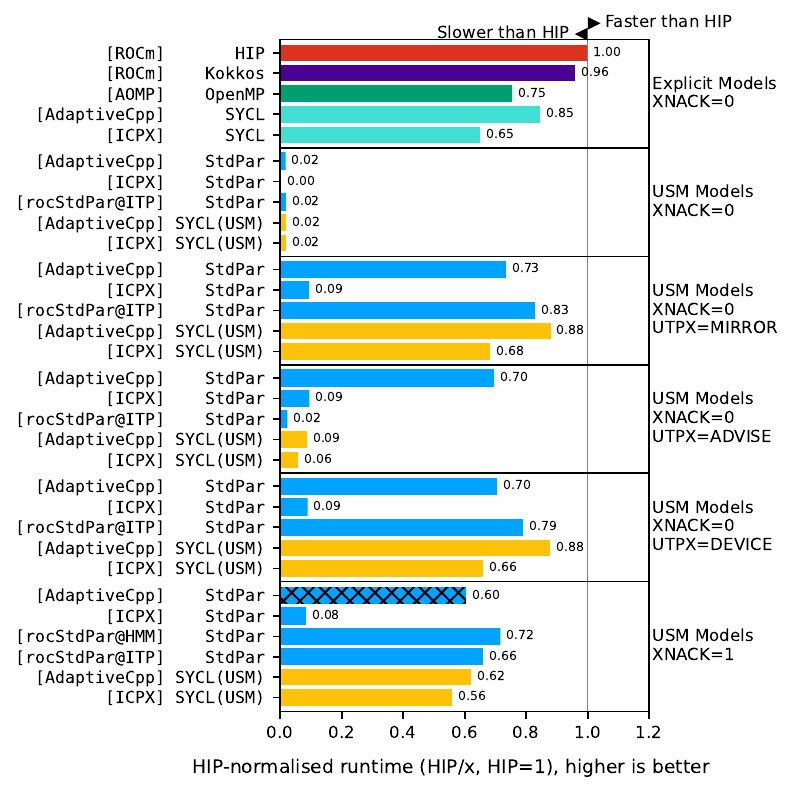}}
    \caption{TeaLeaf normalised runtime across all models against HIP on RadeonVII, \textbf{higher is better}}
    \label{fig:effective-tealeaf-radeonvii}
\end{figure}

\begin{figure}[ht!]
    \centerline{\includegraphics[width=1\linewidth]{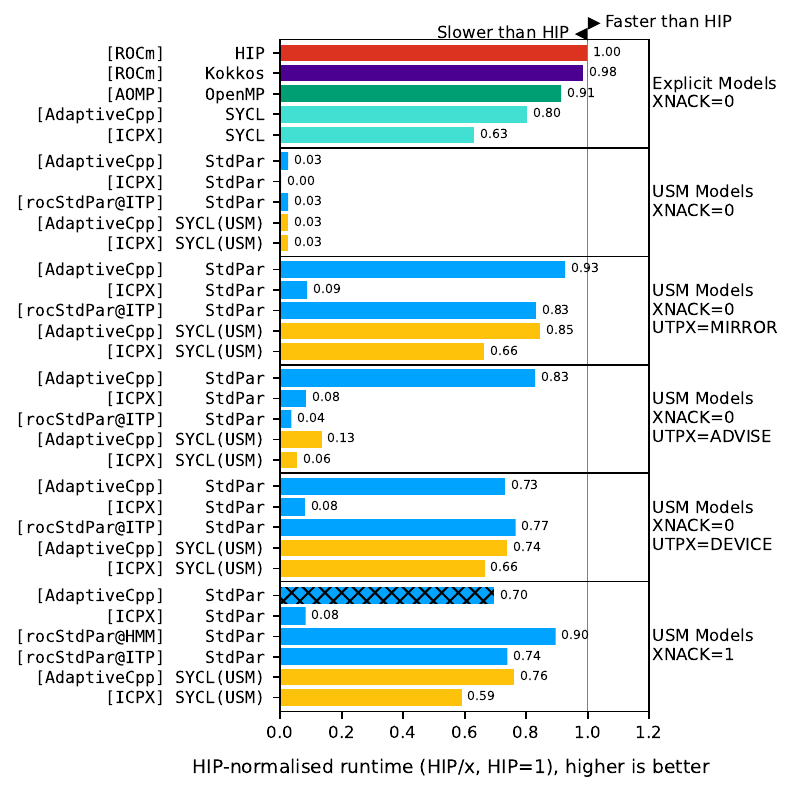}}
    \caption{TeaLeaf normalised runtime across all models against HIP on MI100, \textbf{higher is better}}
    \label{fig:effective-tealeaf-mi100}
\end{figure}

\begin{figure}[ht!]
    \centerline{\includegraphics[width=1\linewidth]{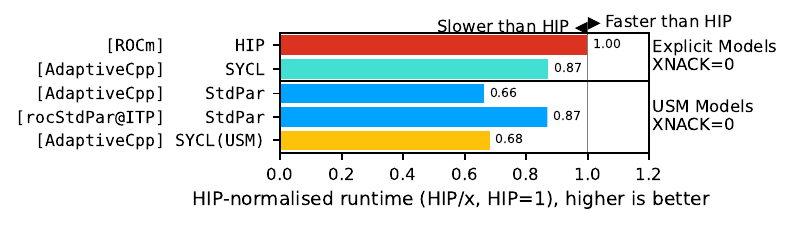}}
    \caption{TeaLeaf normalised runtime across all models against HIP on Raphael, \textbf{higher is better}}
    \label{fig:effective-tealeaf-gfx1036}
\end{figure}

TeaLeaf results are presented in \cref{fig:effective-tealeaf-gfx1036}, \cref{fig:effective-tealeaf-radeonvii}, and \cref{fig:effective-tealeaf-mi100}.
The overall outcome is similar to CloverLeaf discussed in \cref{subsec:results-cloverleaf} with a few notable differences.

Unfortunately, even in an application that has more mixed host-device access, XNACK was still unable to outperform software solutions in a significant way.
Here, UTPX in mirror and device mode both showed performance that is closer to explicit memory models, whereas XNACK's performance was on par with UTPX's mirror mode.

UTPX's advise mode performed very poorly except for AdaptiveCpp\@.
We believe it's the same alignment issue discussed in \cref{subsec:app-babelstream} and \cref{subsec:results-cloverleaf}.

ICPX with oneDPL performed very poorly.
This is due to the suboptimal \texttt{std::transform\_reduce} implementation, possibly because it has not been tuned on AMD platforms.
CloverLeaf did not have this issue because reductions are only used once per time step, whereas two thirds of the core CG solver in TeaLeaf are implemented with iterative calls to \texttt{std::transform\_reduce}.

Results on the Raphael APU are less clear.
While the overall performance for USM models is acceptable, only rocStdPar reached performance parity with explicit models like SYCL\@.
As discussed in \cref{subsec:results-minibude}, AOMP and ICPX do not yet support the Raphael APU\@.

Results with UTPX enabled exhibited the same transient deadlock issue described in \cref{subsec:results-cloverleaf} with a much lower incidence.

    \section{ROCm experience}\label{sec:rocm-experience}
    The overall stability and user experience of the ROCm software stack is poor.
This section details a non-exhaustive list of issues we have encountered while collecting results for this report.

\subsection{Documentation}\label{subsec:documentation}

The HIP API documentation is incomplete.
For example, the recently launched (accessed September 2023) documentation website contains no description at all for the \texttt{hipMalloc} method \footnote{\url{https://docs.amd.com/projects/HIP/en/latest/.doxygen/docBin/html/group___memory.html}}.
Even in the legacy documentation \footnote{\url{https://rocm-developer-tools.github.io/HIP/group__Memory.html}}, the description failed to show a critical difference from CUDA's \texttt{cudaMalloc}: \texttt{hipMalloc} allocates host-accessible (conditional on large-BAR support) but device-resident memory.
In fact, this behaviour is not documented anywhere from AMD; only documentation from ORNL's Crusher\footnote{\url{https://docs.olcf.ornl.gov/systems/crusher_quick_start_guide.html\#enabling-gpu-page-migration}} had details on this.

The lack of detail on critical APIs like these is widespread.
Almost none of the method descriptions in the Managed Memory section\footnote{\url{https://docs.amd.com/projects/HIP/en/latest/.doxygen/docBin/html/group___memory_m.html}} provided any comment on performance, known issues, or expected use cases.
The matching CUDA API documentation does not have this issue.

\subsection{Hardware support}\label{subsec:hardware-support}

ROCm currently only officially supports nine GPU SKUs on Linux.
Of these, RadeonVII is the only consumer-grade card, and it is no longer available on the market for purchase.
As such, it is essentially impossible to validate your code on AMD GPUs without having access to an HPC or cloud service provider.
On NVIDIA platforms, one can simply purchase any consumer GPU and expect near identical baseline software support consistent with that on HPC systems.

Anecdotal evidence suggests that GPUs outside the official ROCm support list will still work.
However, this is not always the case.
For example, AOMP does not currently support first generation Navi GPUs (gfx1012).
We have also observed validation failures with \texttt{hipMallocManaged} on both gfx1012 and gfx1036.
For comparison, the CUDA SDK works on all NVIDIA GPUs (both consumer and HPC) as old as the Kepler generation.

Finally, ROCm 5.7.0 marks gfx906 (RadeonVII and MI50) as EOL, thus eliminating 30\% of the ROCm supported GPUs and also leaves us with no supported consumer GPUs.

\subsection{Validation errors}\label{subsec:validation-errors}

Enabling XNACK on ROCm 5.6.1 gives incorrect results for BabelStream implemented in HIP, rocStdPar, SYCL, and possibly all other models.
This issue is reproducible on RadeonVII\@.
We had brief access to a MI100 with XNACK enabled and were able to verify that this issue is also present on MI100s.
It appears to be a bug in the kernel driver and rolling back to ROCm 5.5.1 or disabling XNACK resolves the issue.

We have confirmation from AMD that this has also been resolved for ROCm 5.7 and newer.

\subsection{Driver quality}\label{subsec:driver-quality}

The KFD kernel driver or the device firmware appears to be quite brittle.
Throughout benchmark runs, we've encountered random kernel panics originating from the amdgpu module, system hangs, and GPU resets due to IB timeout.
On our test system that hosts the RadeonVII (see \cref{tab:software-vers} for the kernel version), when an IB timeout occurs, the kernel initiated BACO reset frequently fails to revive the GPU\@.
Furthermore, a user initiated BACO (Bus-Active, Chip-Off) reset sometimes causes the GPU to go offline for the entire session.

On MI100, we've encountered unexplainable HSA queue hangs which cause the application to deadlock.
The issue only seems to appear during high-load scenarios and is hard to pinpoint.

We have confirmation from AMD that these issues are being investigated, and the situation may improve on Linux kernel 6.7 or newer when paired with an up-to-date ROCm release.

\subsection{Build difficulties}\label{subsec:build-difficulties}

The creation of UTPX is in part motivated by the fact that ROCm is incredibly challenging to build from source.
The difficulty is reflected by the existence of multiple projects \footnote{\url{https://github.com/PawseySC/rocm-from-source}} \footnote{\url{https://github.com/xuhuisheng/rocm-build}} that attempts to document and provide more automated ways of building ROCm.
To add to this, none of the README pages on any of the ROCm's subproject are consistent: many contain outdated build and usage information.

The ROCm binary repositories for package managers do not appear to host source or debug packages.
These are required for gdb to resolve symbols for libraries in the ROCm stack.
Without these packages, gdb is unable to give precise backtraces and shows \texttt{??} next the stack pointer value.
Interestingly, the official documentation that covers HIP debugging \footnote{\url{https://rocm.docs.amd.com/projects/HIP/en/latest/how_to_guides/debugging.html}} provided a sample gdb session that displayed this very issue.

\subsection{Missing tools}\label{subsec:missing-tools}

There is a distinct lack of coherent tooling for AMD GPUs.

\begin{itemize}
    \item ROCm's rocgdb does not work on newer kernel versions such as the one used in our RadeonVII test system, rocgdb reports \texttt{KFD\_IOC\_DBG\_TRAP\_GET\_VERSION}.
    \item ROCm (>= 5.5.1) is released with two profilers, rocprof and rocprofv2, and it's unclear which one is the preferred profiler.
    ROCm appears to have shipped a work-in-progress rocprofv2, as it contains a non-functional plugin system (e.g. the \texttt{att} plugin is incomplete).
    \item Newer profiling tools such as OmniTrace do not support all ROCm platforms: only MI100 and newer CDNA cards are supported.
    Beyond this, there are no working graphical profiling or tracing solutions on Linux as CodeXL is deprecated.
\end{itemize}

    \section{Conclusion}\label{sec:conclusion}

    This report evaluates three emerging StdPar implementations: ROCm StdPar, AdaptiveCpp StdPar, and Intel DPC++ with plugins.
    Overall, for memory-bandwidth bound applications, the performance of all StdPar implementations is tied to the availability and quality of USM, while for compute-bound applications, the performance is highly dependent on compiler optimisations and the programming model.

    We find it encouraging that StdPar is now supported on all major GPU vendors with both first-party and third-party implementations.
    Specifically, AdaptiveCpp and rocStdPar both performed comparably on AMD GPUs, with similar performance to other C++ models like Kokkos or SYCL\@.
    However, we stress that this performance is only achievable with XNACK enabled, or alternatively, with software workarounds such as UTPX\@.
    Unfortunately, DPC++ with oneDPL showed poor performance for memory-bandwidth bound applications, and we suspect the DPC++ vendor plugin for AMD GPU was never tested in production or validated beyond basic smoke tests.

    To provide the USM performance that StdPar requires, we've demonstrated a way to implement USM in userspace without needing compiler/runtime modifications or recompilation of the user program.
    Our UTPX design showed that the fix can be done transparently at the HIP API level; no changes are needed for programming models that use the HIP API, be it SYCL or StdPar.
    We hope that UTPX can motivate AMD to consider adding software-based USM support to improve ROCm's usability and applicability.

    On hardware-accelerated USM, even on GPUs with native support via XNACK, specific, non-default, kernel and software configuration are required.
    The performance of XNACK, in many cases, showed lower performance compared with UTPX\@.
    We find the overall state of USM support on AMD platforms alarming.
    It's unclear how USM will be supported for newer GPUs such as Navi2/3 and onwards due to missing hardware feature.
    Fragmentation of the consumer and HPC space based on feature omissions like this only impacts the applicability of AMD GPUs.

    Finally, our experience compiling this report and past effort porting our mini-apps to AMD GPUs\cite{julia:2021} has highlighted weaknesses in the ROCm software stack.
    These shortcomings encompass documentation, tooling, stability, compatibility, and correctness.
    We hope that AMD continues to make efforts to improve the status quo, allowing programming models like StdPar, and many others, to fully exploit the potential of AMD hardware.

    \section{Future work}\label{sec:future-work}

    StdPar support for AMD platforms is a fast-moving area; as such, the intent of this report is to provide timely feedback for all StdPar implementations.
    To save time, we have omitted experiments with using USM for the HIP model (i.e.\ replacing \texttt{hipMalloc} with \texttt{hipMallocManaged}) as not all mini-apps have this implemented and validated.
    However, enabling this is straightforward and a future study should investigate this.

    Rerunning this study on newer AMD HPC GPUs, such as the MI250X on LUMI, would provide valuable insight on whether XNACK can provide better performance than software-based solutions.
    Specifically, we would like to experiment with AOMP's OpenMP USM support to see how it compares with our results from this report.

    \section{Acknowledgement}\label{sec:acknowledgement}

    We would like to thank Aksel Alpay for his work on AdaptiveCpp and his continuous support on any issues we've raised while compiling this report.

    We would like to thank Tim Dykes' assistance from the HPE HPC/AI EMEA Research Lab, this work is carried out as part of the GW4 Isambard collaboration.

    We would like to thank Alexandru Voicu for his feedback on many of the technical details concerning XNACK, the design of roc-stdpar, and the AMD platform in general.

    This work used results and software developed as part of Intel oneAPI Centre of Excellence.

    This work used the Isambard UK National Tier-2 HPC Service (\url{https://gw4.ac.uk/isambard}) operated by GW4 and the UK Met Office, and funded by EPSRC (EP/P020224/1).
    This work used the DiRAC@Durham facility managed by the Institute for Computational Cosmology on behalf of the STFC DiRAC HPC Facility (www.dirac.ac.uk).
    The equipment was funded by BEIS capital funding via STFC capital grants ST/P002293/1, ST/R002371/1 and ST/S002502/1, Durham University and STFC operations grant ST/R000832/1.
    DiRAC is part of the National e-Infrastructure.
    This work used the HPC Zoo, a multi-platform research cluster managed by the High-Performance Computing Group at the University of Bristol (\url{https://uob-hpc.github.io/zoo}).

        {\footnotesize
    \bibliography{references}
    \bibliographystyle{ieeetr}
    }

    \begin{appendices}
        \crefalias{section}{appendix}
        \section{Artefact description}\label{sec:apdx-index-repro}

Source code for UTPX is available at \url{https://github.com/UoB-HPC/utpx}.
We've used commit \texttt{2a38257b1800e4ac2a2c937ae08e26fd49960ddf} for all experiments involving UTPX\@.
UTPX is built with the default GCC compiler on each platform.
Building UTPX is straightforward and documented in the README of the source code repository.

Source code for BabelStream, with all the models presented in this study, is available at \url{https://github.com/UoB-HPC/BabelStream}.
We've used the \texttt{option\_for\_vec} branch at commit \texttt{87a38e949df2894a7d25ef8782dd96e3978f31ff}.
No modifications were made to BabelStream for any of the benchmarks.

Likewise, source code for miniBUDE is available at \url{https://github.com/UoB-HPC/miniBUDE}.
We've used the \texttt{v2} branch at commit \texttt{bea30762acaefee54ebaf3c68713b66414345e12}.

Finally, source code for CloverLeaf and TeaLeaf is available at \url{https://github.com/UoB-HPC/cloverleaf} and \url{https://github.com/UoB-HPC/tealeaf} respectively.
We've used commit \texttt{4306b008eb21b0dbdad7bd241dfb6a5a337609ca} for CloverLeaf and \texttt{5ee7d753bbb8e8d60b945c0359e00c07aafbab81} for TeaLeaf.

Build flags for each mini-app and compiler combination are recorded at \url{https://github.com/UoB-HPC/performance-portability/tree/2023-benchmarking-amd-stdpar}.
For example, the build flags for miniBUDE on RadeonVII is available under \texttt{benchmarking/2023/bude/radeonvii-local/benchmark.sh}.
    \end{appendices}

\end{document}